\documentstyle[sprocl,epsf]{article}
\def\be{\begin{equation}}
\def\ee{\end{equation}}
\def\bea{\begin{eqnarray}}
\def\eea{\end{eqnarray}}

\def\neb{\hbox{$\overline{\nu}_e \!$ }}

\newcommand{\gapproxeq}{\lower .7ex\hbox{$\;\stackrel{\textstyle >}{\sim}\;$}}
\newcommand{\lapproxeq}{\lower .7ex\hbox{$\;\stackrel{\textstyle <}{\sim}\;$}}

\def\z0{\mbox{$Z^0$}}

\begin{document}

{\hfill DSF 5/99}\vspace*{1cm}

\title{A DETAILED ANALYSIS OF THE NEUTRON TO PROTON RATIO AT THE ONSET OF
PRIMORDIAL NUCLEOSYNTHESIS \footnote{Talk given by S. Esposito at {\it
Second Meeting on New Worlds in Astroparticle Physics}, September 1998,
Faro, Portugal.}} 

\author{S. Esposito, G. Mangano, G. Miele and O. Pisanti}
\address{Dipartimento di Scienze Fisiche and INFN, Sezione di Napoli\\
Mostra d'Oltremare Pad. 20, I-80125, Naples, Italy}

\maketitle 
\abstracts{We report on the results of a recent evaluation of all percent
level corrections to weak rates of processes converting $n \leftrightarrow
p$, which are crucial for the determination of the $^4He$ mass fraction
produced during primordial nucleosynthesis.} 

\section{Introduction}

The more recent data on $^{4}He$ mass fraction and D abundance produced
during Big Bang Nucleosynthesis (BBN) are still controversial, since there
are two different sets of results mutually incompatible \cite{1} 
\begin{eqnarray}
Y_p & = &  0.234 {\pm} 0.0054 ~~~,~~~~~ D/H = (1.9 {\pm} 0.4 ) {\cdot} 10^{-4}~~~;
\nonumber
\\
Y_p &= & 0.243 {\pm} 0.003 ~~~,~~~~~ D/H = (3.40 {\pm} 0.25) {\cdot} 10^{-5}~~~;
\end{eqnarray}
It is possible that forthcoming measurements from high-redshift,
low-metallicity QSO, or better understanding of the present data will
clarify the situation. The fact which however is emerging from the above
results is that Helium data are now reaching a precision of one {\it per
mille}, requiring a similar effort in reducing the theoretical
uncertainties. The $^4He$ mass fraction $Y_p$ crucially depends on the
ratio $n/p$ at the weak interaction freeze--out. To reach a precision of
the order of $10^{-4}$ it is demanding to analyze with an accuracy of one
percent the rates of processes converting $n \leftrightarrow p$, i.e.
$\nu_e ~n \leftrightarrow e^-~ p$, $\neb~p \leftrightarrow e^+ ~n$ and $n
\leftrightarrow e^-~\neb~p$. This means that the simple {\it Born}
approximation rate $\omega_B$, i.e. the tree level $V-A$ interaction in the
{\it infinite mass} limit for nucleons, should be corrected via the
inclusion of several radiative and non radiative effects. We will report on
the results of a recent evaluation of all the above corrections \cite{2}.
For an extensive discussion, as well as a detailed analysis of the
calculation, see reference 2. 

\section{Corrections to Born rates and $^4He$ mass fraction}

Leading corrections to the weak rates of neutron decay and electron and
neutrino capture processes can be classified as follows: 
\begin{itemize} 

\item[i)] order $\alpha$ radiative corrections $\Delta \omega_R$. These
effects have been extensively studied in literature and can be classified
in {\it outer} factors, involving the nucleon as a whole, and {\it inner }
ones, which instead depend on the details of nucleon internal structure.
Other small effects are actually expected at higher order in $\alpha$,
since, for example, the estimated value of the neutron lifetime is
compatible with the experimental value $\tau_n^{exp}$ at 4-$\sigma$ level
only. These additional contributions are usually taken into account by
eliminating the coupling in front of the reaction rates in favour of
$\tau_n^{exp}$. In Figure 1 we report the result of our calculation
\cite{2} for $\Delta \omega_R$ as a function of the ratio $T/m_e$. 

\item[ii)] All Born amplitudes should be also corrected for nucleon finite
mass effects $\Delta \omega_K$. They affect the allowed phase space as well
as the weak amplitudes, which should now include the contribution of
nucleon weak magnetism. Initial nucleons with finite mass will also have a
thermal distribution in the comoving frame, over which it is necessary to
average the transition probabilities. All these effects are of the order of
$T/M_N$ or $m_e/M_N$, with $M_N$ the nucleon mass. The percent corrections
$\Delta \omega_K/\omega_B$ are reported in Figure 2. 

\item[iii)] Since all reactions take place in a thermal bath of electron,
positron, neutrinos, antineutrinos and photons, finite density and
temperature radiative effects $\Delta \omega_T$ should be also included;
these corrections, of the order of $\alpha (T/m_e)$, account for the
electromagnetic interactions of the in/out particles, involved in the
microscopic weak processes $n\leftrightarrow p$, with the surrounding
medium. They can be evaluated in the real time formalism for finite
temperature field theory \cite{2}. As can be seen in Figure 3, these
corrections only represents $0.2 \div 0.3 \%$ of the corresponding Born
rates at the freeze out temperature $T \sim 1 MeV$. 

\end{itemize}
The corrections to the $n \leftrightarrow p$ rates may produce a sensible
effect on the $^4He$ mass fraction $Y_p$, which is strongly dependent on
the neutron fraction $X_n = n_n/(n_n+n_p)$ at the nucleon freeze-out. The
expected variation of the surviving neutron fraction $\delta X_n$ induced
by the whole effects $\Delta \omega$, which we report in Figure 4, can be
evaluated by solving the corresponding transport equation. We find \cite{2}
for the asymptotic abundance $\delta X_n\simeq 0.0024$, with a relative
change, in percent, $\delta X_n/X_n = 1.6 \%$. These results allows for a
simple estimation of the corrections to $^4He$ mass fraction $\delta Y_p =
2 ~\delta X_n~ \exp \left(- {t_{ns}}/{\tau_n^{exp}} \right)$, where $t_{ns}
\simeq 180~ sec$ corresponds to the onset of nucleosynthesis and the
exponential factor accounts for the depletion of relic neutrons at
freeze-out due to $\beta$-decay. Using the results for $\delta X_n$ we find
$\delta Y_p  \simeq  0.004$, or, in percent ${\delta Y_p}/{Y_p} = 1.6 \%$.

\newpage
\begin{figure}
\epsfxsize=7cm
\epsfysize=7cm
\epsffile{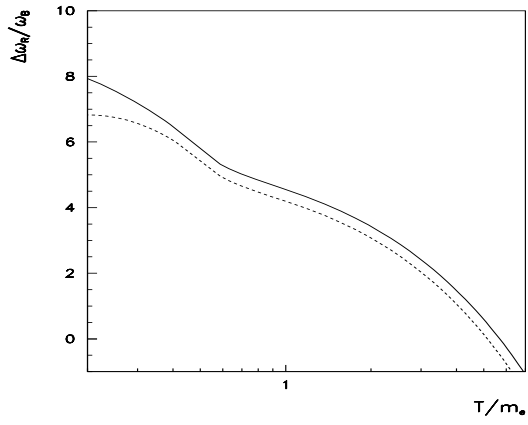}
\caption{{\footnotesize Zero temperature radiative corrections, for $n
\rightarrow p$ (solid line) and $p \rightarrow n$ (dashed line) processes,
in percent.}} 
\end{figure}
\begin{figure}
\epsfxsize=7cm
\epsfysize=7cm
\epsffile{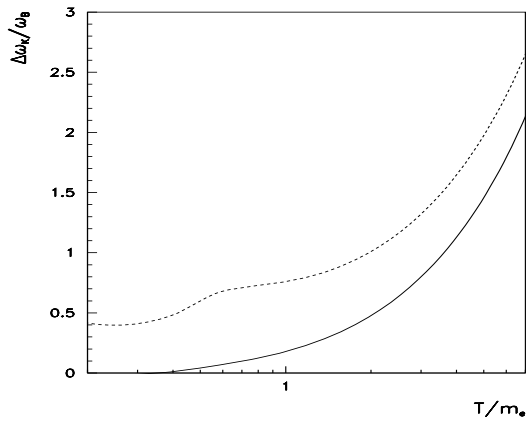}
\caption{{\footnotesize  The finite nucleon mass contributions  to the $n
\rightarrow p$ (solid line) and $p \rightarrow n$ (dashed line)
processes.}} 
\end{figure}
\begin{figure}
\epsfxsize=7cm
\epsfysize=7cm
\epsffile{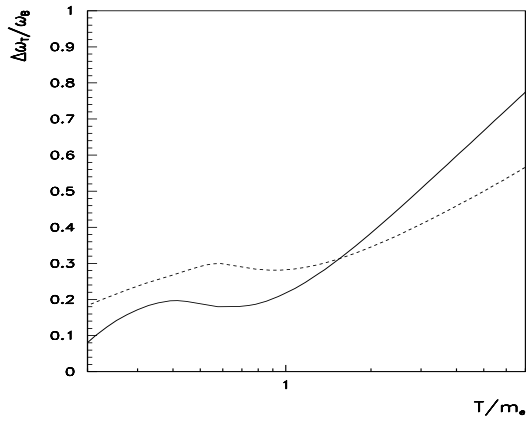}
\caption{{\footnotesize The thermal radiative corrections to the $n
\rightarrow p$ (solid line) and $p \rightarrow n$ (dashed line) processes,
expressed in percent.}} 
\end{figure}
\begin{figure}
\epsfxsize=7cm
\epsfysize=7cm
\epsffile{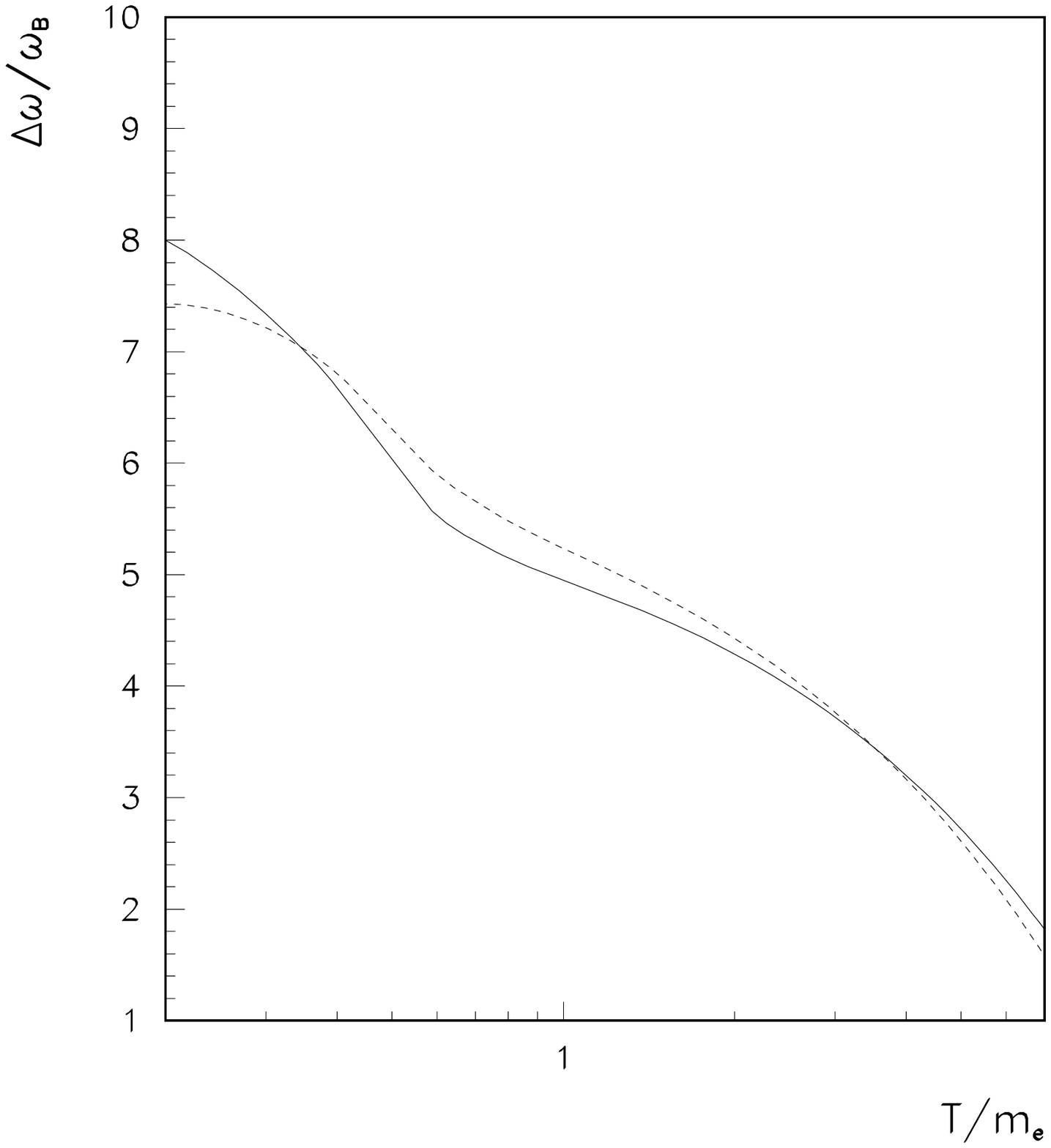}
\caption{{\footnotesize The total relative corrections $\Delta \omega
/\omega_B$ for the $n \rightarrow p$ (solid line) and $p \rightarrow n$
(dashed line) processes, expressed in percent.}} 
\end{figure}


\section*{References}

\begin{thebibliography}{99}

\bibitem{1} K.A. Olive and D. Thomas, {\it Astropart. Phys.} {\bf 7}, 27
(1997);
 Y.I. Izotov, T.X. Thuan and V.A. Lipovetsky, {\it Ap. J. Suppl.} {\bf
 108}, 1 (1997).

\bibitem{2} S. Esposito, G. Mangano, G. Miele and O. Pisanti,
{\it Nucl. Phys.} B {\bf 5842}, 1 (1998).

\end{thebibliography}
\end{document}